\documentclass{ws-ijbc}

\usepackage{graphicx}
\graphicspath{{figures_pdf/}}
\usepackage{multirow,bigstrut}
\usepackage{amssymb,amsmath,bm}

\usepackage[normalem]{ulem}

\usepackage{xcolor}

\newlength\figwidth
\usepackage{color}
\definecolor{dgreen}{rgb}{0,.6,0}

\begin{document}

\catchline{}{}{}{}{} 

\markboth{C. Li et al.}{Breaking an image encryption algorithm}

\title{Breaking an image encryption algorithm based on chaos}

\author{Chengqing Li\textsuperscript{1,2}\thanks{Corresponding author. Email: DrChengqingLi@gmail.com.},
Michael Z. Q. Chen\textsuperscript{3}, and Kwok-Tung Lo\textsuperscript{1}\\
\textsuperscript{1} College of Information Engineering,\\ Xiangtan University, Xiangtan 411105, Hunan, China\\[0.5em]
\textsuperscript{2} Department of Electronic and Information Engineering, \\The Hong Kong Polytechnic University, Hong Kong\\[0.5em]
\textsuperscript{3} Department of Mechanical Engineering,\\ The University of Hong Kong, Hong Kong}

\maketitle

\begin{history}
Oct 25, 2010
\end{history}

\begin{abstract}
Recently, a chaos-based image encryption algorithm called MCKBA (Modified Chaotic-Key Based Algorithm) was proposed.
This paper analyzes the security of MCKBA and finds that it can be broken with a differential attack, which requires only four chosen plain-images. Performance of the attack is verified by experimental results. In addition, some defects of MCKBA,
including insensitivity with respect to changes of plain-image/secret key, are reported.
\end{abstract}


\section{Introduction}
\noindent

Rapid development of information technology and popularization of digital products
require that multimedia data are transmitted over all kinds of wired/wireless
networks more and more frequently. Therefore, secure delivery of multimedia data becomes increasingly important. However,  traditional text encryption schemes fail to be competent for the task due to the big differences between textual and multimedia data.
Under the pressure of this challenge, researchers attempted to propose special multimedia encryption schemes
utilizing all kinds of nonlinear theories in the past decade. The subtle similarity between chaos and
cryptography makes chaos considered as an ideal tool to design secure and efficient encryption
schemes and a great number of
multimedia encryption schemes based on it have been presented
\cite{Chen&Yen:RCES:JSA2003,YaobinMao:CSF2004,Flores:EncryptLatticeChaos06,Xiang:Encrypt:PLA07,
Ye:Scramble:PRL10,KWong:chaoticoding:TCASSII10}.  Unfortunately, many of them have been found to be
insecure and/or incomplete from the viewpoint of modern cryptology
\cite{Kaiwang:PLA2005,Li:AttackingRCES2008,Li:AttackingPOMC2008,David:AttackingChaos08,Rhouma:BreakLian:PLA08,
Zhou:CommentChaoticrypt:TCASI08,Solak:BreakXiang:PLA09,
Li:BreakImageCipher:IVC09,Li:AttackingIVC2009,Solak:Fridrich:IJBC10,YangXiao:4RBlock:CNSNS10}. References \cite{AlvarezLi:Rules:IJBC2006,
Li:ChaosImageVideoEncryption:Handbook2004} conclude some general rules about
evaluating the security of chaos-based encryption schemes.

In \cite{Yen:CKBA:ISCAS2000}, a chaotic key-based algorithm (CKBA) for image
encryption was proposed. The algorithm encrypts each pixel by four possible operations: XORing or XNORing
it with one of two predefined sub-keys. A pseudo-random number sequence (PRNS), obtained from a one-dimensional chaotic system, is
used to determine which operation is exerted. As shown in \cite{Li-Zheng:CKBA:ISCAS2002}, CKBA can be easily
broken with only one known/chosen-image. To enhance security of CKBA against known/chosen-plaintext attack,
\cite{Rao:ModifiedCKBA:ICDSP07} proposes a modified chaotic-key based algorithm (MCKBA) by employing a modular addition operation like \cite{SocekLi:SecureComm2005}. To further enhance the security against brute-force attack, \cite{CH:HCKBA:IJBC10} replaces
the one-dimensional chaotic system generating PRNS with a simple hyperchaos generator proposed in \cite{Takahashi:HyperChaos:TCASII04} and
names the algorithm HCKBA (Hyper Chaotic-Key Based Algorithm). Since the two schemes MCKBA and HCKBA share the same
structure, this paper only analyzes the security of MCKBA and finds that the scheme can be broken with only four chosen plain-images.
Both theoretical analysis and experimental results are provided to support the conclusion. In addition, some other security defects of MCKBA,
including insensitivity with respect to changes of plain-image/secret key, are discussed.

The rest of this paper is organized as follows. The image encryption algorithm under study is
introduced in Sec.~\ref{sec:scheme}. Detailed cryptanalysis on the algorithm is presented in
Sec.~\ref{sec:cryptanalysis} with experimental results. The last section concludes this paper.

\section{Modified Chaotic-Key Based Algorithm (MCKBA)}
\label{sec:scheme}

The plaintext encrypted by MCKBA is a gray-scale image of size $M\times
N$ (width$\times$height). The plain-image is scanned in the raster order and represented as a 1D signal
$\bm{I}=\{I(i)\}_{i=0}^{MN-1}$. Then, a binary sequence $\bm{I}_b=\{I_b(l)\}_{l=0}^{8MN-1}$ is constructed,
where $\sum_{j=0}^7I_b(8\cdot i+j)\cdot 2^j=I(i)$ $\forall\ i\in \{0,\cdots, MN-1\}$. With a pre-defined integer parameter $n$, an $n$-bit number sequence $\bm{J}=\{J(i)\}_{i=0}^{\lceil 8MN/n\rceil-1}$ is generated for encryption, where $J(i)=\sum_{j=0}^{n-1}I_b(n\cdot i+j)\cdot 2^j$. Note that sequence $\bm{I}_b$ is padded with some
zero bits if $(8MN)$ is not a multiple of $n$. Without loss of generality, assume $n$ can divide $(8MN)$ here. MCKBA operate on the intermediate sequence $\bm{J}$ and get
$\bm{J}'=\{J'(i)\}_{i=0}^{8MN/n-1}$, where $J'(i)=\sum_{j=0}^{n-1}I_b'(n\cdot i+j)\cdot 2^j$. Finally, cipher-image
$\bm{I}'=\{I'(i)\}_{i=0}^{MN-1}$ is obtained, where $I'(i)=\sum_{j=0}^7I_b'(8\cdot i+j)\cdot 2^j$. With the above notations, MCKBA can be described
as follows\footnote{To make the presentation more concise and consistent, some notations in the original paper \cite{Rao:ModifiedCKBA:ICDSP07}
are modified, and some details of MCKBA are also supplied.}.

\begin{itemlist}
\item \textit{The secret key}: two random numbers $key_1$, $key_2\in\{0, \cdots, 2^n-1\}$, and the initial condition $x(0)\in(0,1)$
of the following chaotic Logistic map:
\begin{equation}
x(i+1)=3.9\cdot x(i)\cdot(1-x(i)),\label{equation:Logistic}
\end{equation}
where $\sum_{j=0}^{n-1}(key_{1,j}\oplus key_{2,j})=\lceil n/2\rceil$, $key_1=\sum_{j=0}^{n-1}key_{1,j}\cdot 2^j$, $key_2=\sum_{j=0}^{n-1}key_{2,j}\cdot 2^j$, and
$\oplus$ denotes eXclusive OR (XOR) operation.

\item \textit{Initialization}: run the chaotic system to generate a chaotic sequence,
$\{x(i)\}_{i=0}^{MN/(2n)-1}$. From the 32-bit binary representation of $x(i)=\sum_{j=1}^{32}b(32\cdot i+j-1)\cdot 2^{-j}$,
derive a pseudo-random binary sequence (PRBS), $\{b(l)\}_{l=0}^{16MN/n-1}$.

\item \textit{Encryption}: for the $i$-th plain-element $J(i)$, $i=0\sim 8MN/n-1$, the corresponding cipher-element
$J'(i)$ is determined by the following rule:
\begin{equation}
J'(i)=
\begin{cases}
(J(i)\dotplus key_1)\oplus key_1, & \mbox{if }B(i)=3,\\
(J(i)\dotplus key_1)\odot  key_1, & \mbox{if }B(i)=2,\\
(J(i)\dotplus key_2)\oplus key_2, & \mbox{if }B(i)=1,\\
(J(i)\dotplus key_2)\odot  key_2, & \mbox{if }B(i)=0,
\end{cases}
\label{eq:Encrypt}
\end{equation}
where $B(i)=2\cdot b(2i)+b(2i+1)$, $a\dotplus b=(a+b)\bmod 2^n$ and
$\odot$ denotes XNOR operation. Since $a\odot b=\overline{a\oplus b}=a\oplus\bar{b}$, the above
equation is equivalent to
\begin{equation}
J'(i)=
\begin{cases}
(J(i)\dotplus key_1)\oplus key_1,             & \mbox{if }B(i)=3,\\
(J(i)\dotplus key_1)\oplus \overline{key_1},  & \mbox{if }B(i)=2,\\
(J(i)\dotplus key_2)\oplus key_2,             & \mbox{if }B(i)=1,\\
(J(i)\dotplus key_2)\oplus \overline{key_2},  & \mbox{if }B(i)=0.
\end{cases}
\label{eq:equiEncrypt}
\end{equation}

\item \textit{Decryption}: the decryption procedure is similar to that of the encryption, but
with Eq.~(\ref{eq:equiEncrypt}) replaced by following
\begin{equation}
J(i)=
\begin{cases}
(J'(i)\oplus key_1)\dot{-}key_1,            & \mbox{if }B(i)=3,\\
(J'(i)\oplus \overline{key_1})\dot{-}key_1, & \mbox{if }B(i)=2,\\
(J'(i)\oplus key_2)\dot{-}key_2,            & \mbox{if }B(i)=1,\\
(J'(i)\oplus \overline{key_2})\dot{-}key_2, & \mbox{if }B(i)=0,
\end{cases}
\end{equation}
where $a\dot{-} b=(a-b+2^n)\bmod 2^n$.
\end{itemlist}

\section{Cryptanalysis}
\label{sec:cryptanalysis}

\subsection{The Differential Attack}

Differential attack is usually a chosen-plaintext attack, assuming that the attacker can obtain cipertexts for some set of
chosen plaintexts. The goal of the attack is to gain information about the secret key or plaintext by analyzing how differences
in the chosen plaintexts affect the resultant difference at the corresponding ciphertexts. Note that difference is defined with
respect to any given operation, e.g., XOR. In \cite[III.B]{Rao:ModifiedCKBA:ICDSP07} and \cite[Sec.~3.2]{CH:HCKBA:IJBC10},
the authors claimed that MCKBA is very robust against chosen-plaintext attack. However, we will show how it can be broken very
easily with only four chosen plain-images.

Since plain-image and intermediate sequences $\bm{J}$ can be obtained from each other without any secret key, choosing the former is actually equivalent to
choosing the latter. If two known intermediate sequences $\bm{J}_1=\{J_1(i)\}_{i=0}^{8MN/n-1}$ and $\bm{J}_2=\{J_2(i)\}_{i=0}^{8MN/n-1}$ are
encrypted with the same secret key, their corresponding encrypted results $\bm{J}'_1=\{J_1'(i)\}_{i=0}^{8MN/n-1}$ and $\bm{J}'_2=\{J_2'(i)\}_{i=0}^{8MN/n-1}$
satisfy the following relation
\begin{equation}
J_1'(i)\oplus J_2'(i)=
\begin{cases}
(J_1(i)\dotplus key_1)\oplus (J_2(i)\dotplus key_1),& \mbox{if }B(i)\in\{2, 3\},\\
(J_1(i)\dotplus key_2)\oplus (J_2(i)\dotplus key_2),& \mbox{if }B(i)\in\{0, 1\}.
\end{cases}
\label{eq:DifferentialEquation}
\end{equation}

Regardless the value of $B(i)$, $(J_1'(i)\oplus J_2'(i))$ can be represented by an equation
in the following form
\begin{equation}
y=(a\dotplus x)\oplus (b\dotplus x),
\label{eq:encryptform}
\end{equation}
where $a, b, x, y\in\{0, \cdots, 2^n-1\}$.

The following theorem discusses how to solve the above equation.
\begin{theorem}
Assume that $a, b, x$ are all $n$-bit integers, then a lower bound on the number of queries $(a, b)$ to
solve Eq.~(\ref{eq:encryptform}) for any $x$ is (i) 0 if $n=1$; (ii) 1 if $n=2$;
(iii) 2 if $n=3$; or (iv) 3 if $n\ge 4$.
\label{the:num}
\end{theorem}
\begin{proof}
First, rewrite Eq.~(\ref{eq:encryptform}) as the following equivalent form
\begin{equation}
\tilde{y}=y\oplus a\oplus b=(a\dotplus x)\oplus (b\dotplus x)\oplus a\oplus b.
\label{eq:encryptform2}
\end{equation}
Let $x=\sum_{j=0}^{n-1}x_j\cdot 2^j$, $a=\sum_{j=0}^{n-1}a_j\cdot 2^j$, $b=\sum_{j=0}^{n-1}b_j\cdot 2^j$, and
$\tilde{y}=\sum_{j=0}^{n-1}\tilde{y}_j\cdot 2^j$. Then, except $\tilde{y}_0\equiv 0$, Eq.~(\ref{eq:encryptform2}) can be decomposed into
the following iteration form
\begin{equation}
\left\{
\begin{array}{ccl}
c_{i+1}         & = & (x_i\cdot a_i) \oplus (x_i\cdot c_i) \oplus (a_i\cdot c_i),\\
\tilde{c}_{i+1} & = & (x_i\cdot b_i) \oplus (x_i\cdot \tilde{c}_i) \oplus (b_i\cdot \tilde{c}_i),\\
\tilde{y}_{i+1} & = & c_{i+1}\oplus \tilde{c}_{i+1},
\end{array}\right.
\label{eq:bitdecomposition}
\end{equation}
where $i\in \{0, \cdots, n-2\}$, $c_0=0$, $\tilde{c}_{0}=0$.

Table~1 lists the values of $\tilde{y}_{i+1}$ under all possible different values of $a_i, b_i,\tilde{y}_{i}, x_i, c_i$. From Table~1, one can see that the values of unknown bit $x_i$ can be determined if and only if $(a_i, b_i,\tilde{y}_{i} )$ falls in the 1, 2, 4, 7-th column (zero-based) of the table, namely
\begin{equation}
(a_i+b_i\cdot 2+\tilde{y}_{i}\cdot 2^2)\in \{1,2,4,7\}.
\label{eq:condition}
\end{equation}
\begin{table}[htbp]
\tbl{The values of $\tilde{y}_{i+1}$ corresponding to the values of $a_i, b_i,\tilde{y}_{i}, x_i, c_i$.}
{\begin{tabular}{*{8}{c}c}
\toprule
\multirow{2}{0.3in}{$(x_i, c_i)$}     & \multicolumn{8}{c}{$(a_i, b_i,\tilde{y}_{i} )$} \\
\cline{2-9} & $(0,0,0)$ & $(0,0,1)$ & $(0,1,0)$ & $(0,1,1)$ & $(1,0,0)$ & $(1,0,1)$ & $(1,1,0)$ & $(1,1,1)$\\\hline
(0, 0)      &     0     &   0       &   0       &   1       &   0       &   0       &       0   &   1    \\
(0, 1)      &     0     &   0       &   1       &   0       &   1       &   1       &       0   &   1    \\  \hline
(1, 0)      &     0     &   1       &   1       &   1       &   1       &   0       &       0   &   0    \\
(1, 1)      &     0     &   1       &   0       &   0       &   0       &   1       &       0   &   0    \\ [-4pt]
\botrule
\end{tabular}}
\label{table:carrybits}
\end{table}

When $n=1$, Eq.~(\ref{eq:encryptform2}) becomes $\tilde{y}\equiv 0$. So, no pair of $(a, b)$ is required to achieve
the value of $x$. Since $\tilde{y}_{n-1}$ bears no relation with $x_{n-1}$, we only need to discuss how to
obtain the $(n-1)$ least significant bits of $x$ for other values of $n$.
\begin{itemlist}
\item $n=2$: Since $\tilde{y}_0=0$, $c_0=0$, one can get $x_0=\tilde{y}_1$ by setting $(a_0, b_0)=(1, 0)$;

\item $n=3$: No matter what $(a_0, b_0)$ is, $y_1\in\{0, 1\}$. Therefore, it is impossible to obtain $x_1$ with only set of $(a_1, b_1)$ for any $x$.
Select $(a, b)$ satisfying that $(a_0, b_0)=(1, 0)$, get $x_0$ as the above case. Let $(a_1, b_1)=(1, 0)$, $x_1$ can be determined if $y_1=0$; otherwise we have to resort to another query $(a', b')$. Let $\tilde{y}'=\sum_{j=0}^{n-1}\tilde{y}'_j\cdot 2^j$ denote the output of Eq.~(\ref{eq:encryptform2}) corresponding to the second query. Set $(a'_0, b'_0)=(a_0, b_0)$ and $(a'_1, b'_1)=(1,1)$
if $y_1=1$, then we can get $x_1=\overline{y_2'}$;

\item $n\ge 4$: In this case, $(\tilde{y}_1,\tilde{y}_2)$ and $(\tilde{y}'_1,\tilde{y}_2')$ can be all possible values. Observing Table~1,
it can be easily verified that there is no $(a,b)$ and $(a',b')$ satisfying either Eq.~(\ref{eq:condition}) or
\begin{equation}
(a_i'+b_i'\cdot 2+\tilde{y}_{i}'\cdot 2^2)\in \{1,2,4,7\}
\end{equation}
for $i=1, 2$. This means $x_2$ cannot  always be determined. Therefore, we need one more query $(a^{\star}=\sum_{j=0}^{n-1}a^{\star}_j\cdot 2^j, b^{\star}=\sum_{j=0}^{n-1}b^{\star}_j\cdot 2^j)$.
Let $\tilde{y}^{\star}=\sum_{j=0}^{n-1}\tilde{y}^{\star}_j\cdot 2^j$ denote the corresponding output with respect to Eq.~(\ref{eq:encryptform2}).
Given a set of $(a_{i+k},b_{i+k},a'_{i+k},b'_{i+k},a^{\star}_{i+k},b^{\star}_{i+k})$, one can get $(c_{i+k+1}, \tilde{y}_{i+k+1}, c'_{i+k+1}, \tilde{y}'_{i+k+1}, c^{\star}_{i+k+1}, \tilde{y}^{\star}_{i+k+1})$ from $(c_{i+k}, \tilde{y}_{i+k}, c'_{i+k}, \tilde{y}'_{i+k}, c^{\star}_{i+k}, \tilde{y}^{\star}_{i+k})$ and value of $x_{i+k}$, where $i, k\in\mathbb{Z}$. Let arrows of plain head and ``V-back" head denote $x_{i+k}=0$ and $x_{i+k}=1$ respectively, Figure~\ref{fig:RelationBlock} illustrates mapping relationship between $(c_{i+k}, \tilde{y}_{i+k}, c'_{i+k}, \tilde{y}'_{i+k}, c^{\star}_{i+k}, \tilde{y}^{\star}_{i+k})$ and
$(c_{i+k+1}, \tilde{y}_{i+k+1}, c'_{i+k+1}, \tilde{y}'_{i+k+1}, c^{\star}_{i+k+1}, \tilde{y}^{\star}_{i+k+1})$ for a given $(a_{i+k},b_{i+k},a'_{i+k},b'_{i+k},a^{\star}_{i+k},b^{\star}_{i+k})$, where $k=0, 1, 2$. Since
$(c_0,\tilde{y}_{0},c'_0,\tilde{y}'_{0},c^{\star}_0,\tilde{y}^{\star}_{0})\equiv (0, 0, 0, 0, 0, 0)$, the dashed arrows in Fig.~\ref{fig:RelationBlock} describe Eq.~(\ref{eq:bitdecomposition}) with the three sets of $(a, b)$ for $i=0, 1, 2$. Note that the data in the fourth column of the table shown
in Fig.~\ref{fig:RelationBlock} is exactly the same as the first one. Therefore, Fig.~\ref{fig:RelationBlock} shows calculation of Eq.~(\ref{eq:encryptform2})
under all different bit levels if the variable $i$ shown in Fig.~\ref{fig:RelationBlock} go through $3\cdot t$, where $t=0\sim \lfloor n/3\rfloor$ and $i+k\leq n-1$. From Fig.~\ref{fig:RelationBlock}, it can be easily verified that the following relationship
\begin{equation}
\left(a_{i+k}+b_{i+k}\cdot 2+\tilde{y}_{i+k}\cdot 2^2, a_{i+k}'+b_{i+k}'\cdot 2+\tilde{y}_{i+k}'\cdot 2^2, a^{\star}_{i+k}+b^{\star}_{i+k}\cdot 2+\tilde{y}^{\star}_{i+k}\cdot 2^2 \right)\cap\{1, 2, 4, 7\} \neq \emptyset
\end{equation}
is always satisfied, which means $x_{i+k}$ can be derived from Table~\ref{table:carrybits}. This completes the proof.

\begin{figure}[!htb]
\centering
\begin{minipage}{1.8\figwidth}
\centering
\includegraphics[width=\textwidth]{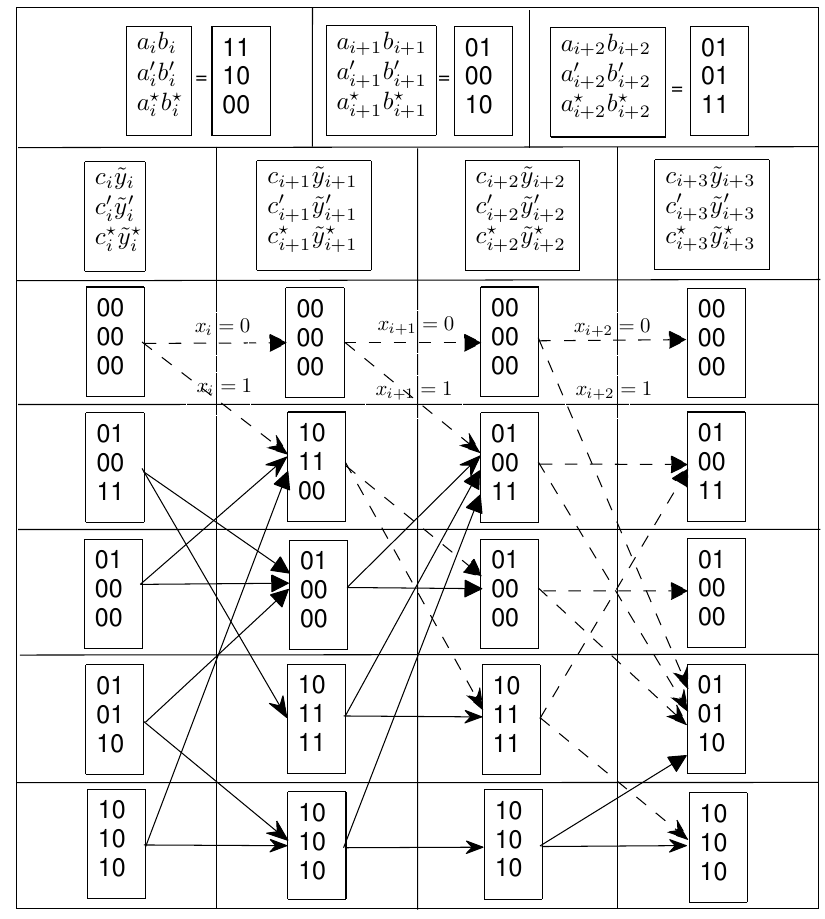}
\end{minipage}
\caption{Relationship between $(c_{i+k}, \tilde{y}_{i+k}, c'_{i+k}, \tilde{y}'_{i+k}, c^{\star}_{i+k}, \tilde{y}^{\star}_{i+k})$ and
$(c_{i+k+1}, \tilde{y}_{i+k+1}, c'_{i+k+1}, \tilde{y}'_{i+k+1}, c^{\star}_{i+k+1}, \tilde{y}^{\star}_{i+k+1})$ for a given $(a_{i+k},b_{i+k},a'_{i+k},b'_{i+k},a^{\star}_{i+k},b^{\star}_{i+k})$, where
$k=0, 1, 2$.}
\label{fig:RelationBlock}
\end{figure}
\end{itemlist}
\end{proof}

\begin{corollary}
The $(n-1)$ least significant bits of $x$ in Eq.~(\ref{eq:encryptform}) can be determined easily
by setting $(a, b)$ with the following three sets of numbers
\label{coro:msb}
\begin{eqnarray*}
\left\{ \left(\sum\nolimits_{j=0}^{\lceil n/3\rceil-1}(100)_2\cdot 8^j\right)\bmod 2^n, \left(\sum\nolimits_{j=0}^{\lceil n/3\rceil-1}(111)_2\cdot 8^j\right)\bmod 2^n\right\};\\
\left\{ \left(\sum\nolimits_{j=0}^{\lceil n/3\rceil-1}(100)_2\cdot 8^j\right)\bmod 2^n, \left(\sum\nolimits_{j=0}^{\lceil n/3\rceil-1}(001)_2\cdot 8^j\right)\bmod 2^n\right\};\\
\left\{ \left(\sum\nolimits_{j=0}^{\lceil n/3\rceil-1}(011)_2\cdot 8^j\right)\bmod 2^n, \left(\sum\nolimits_{j=0}^{\lceil n/3\rceil-1}(001)_2\cdot 8^j\right)\bmod 2^n\right\}{\ }
\end{eqnarray*}
and checking the corresponding $\tilde{y}=y\oplus a\oplus b$.
\end{corollary}
\begin{proof}
The proof is straightforward.
\end{proof}

\begin{proposition}
Assume that $a$ and $x$ are both $n$-bit integers, $n\in \mathbb{Z}^+$, one has the following two equations
\begin{eqnarray}
(a\oplus x)\dot{-}x            & = & (a\oplus x\oplus 2^{n-1})\dot{-}(x\oplus 2^{n-1}),           \label{eq:EquiKey1}\\
(a\oplus \overline{x})\dot{-}x & = & (a\oplus \overline{x\oplus 2^{n-1}})\dot{-}(x\oplus 2^{n-1}).\label{eq:EquiKey2}
\end{eqnarray}
\label{propsition:msb}
\end{proposition}
\begin{proof}
Eq.~(\ref{eq:EquiKey1}) can be proved under four conditions.
i) When $(a\oplus x)\ge 2^{n-1}$ and $x\ge 2^{n-1}$:
$(a\oplus x\oplus 2^{n-1})\dot{-}(x\oplus 2^{n-1})=((a\oplus x)-2^{n-1})\dot{-}(x-2^{n-1})=(((a\oplus x)-2^{n-1})-(x-2^{n-1})+2^n)\bmod 2^n=(a\oplus x)\dot{-}x$;
ii) When $(a\oplus x)\ge 2^{n-1}$ and $x<2^{n-1}$: $(a\oplus x\oplus 2^{n-1})\dot{-}(x\oplus 2^{n-1})=((a\oplus x)-2^{n-1})\dot{-}(x+2^{n-1})=(((a\oplus x)-2^{n-1})-(x+2^{n-1})+2^n)\bmod 2^n=(a\oplus x)\dot{-}x$;
iii) When $(a\oplus x)<2^{n-1}$ and $x\ge 2^{n-1}$: $(a\oplus x\oplus 2^{n-1})\dot{-}(x\oplus 2^{n-1})=((a\oplus x)+2^{n-1})\dot{-}(x-2^{n-1})=(a\oplus x)\dot{-}x$;
iv) When $(a\oplus x)<2^{n-1}$ and $x<2^{n-1}$: $(a\oplus x\oplus 2^{n-1})\dot{-}(x\oplus 2^{n-1})=((a\oplus x)+2^{n-1})\dot{-}(x+2^{n-1})=(a\oplus x)\dot{-}x$.
Similarly, Eq.~(\ref{eq:EquiKey2}) can be proved.
\end{proof}

Corollary~\ref{coro:msb} means that one can only choose four intermediate sequences, $\bm{J}_0$, $\bm{J}_1$, $\bm{J}_2$ and $\bm{J}_3$, to break MCKBA, where
\begin{eqnarray}
J_0(i) & \equiv &(\sum\nolimits_{j=0}^{\lceil n/3\rceil-1}1\cdot 8^j)\bmod 2^n,\nonumber\\
J_1(i) & \equiv &(\sum\nolimits_{j=0}^{\lceil n/3\rceil-1}7\cdot 8^j)\bmod 2^n,\label{eq:scondchooseimage}\\
J_2(i) & \equiv &(\sum\nolimits_{j=0}^{\lceil n/3\rceil-1}4\cdot 8^j)\bmod 2^n,\nonumber\\
J_3(i) & \equiv &(\sum\nolimits_{j=0}^{\lceil n/3\rceil-1}6\cdot 8^j)\bmod 2^n\nonumber.
\end{eqnarray}
With respect to the 1D representation of 2D images defined in Sec.~\ref{sec:scheme},
basic repeated pattern of the corresponding gray-scale images of $\bm{J}_0$, $\bm{J}_1$, $\bm{J}_2$ and $\bm{J}_3$ are $[73, 146, 36]$, $[255, 255, 255]$, $[36, 73, 146]$, $[182, 109, 219]$, respectively.
As shown in Proposition~\ref{propsition:msb}, the unknown most significant bits of $key_1$ and/or $key_2$ have no influence on decryption of MCKBA, so they are
considered being recovered correctly in rest of the paper. Let $key^*(i)$ denote solution of Eq.~(\ref{eq:DifferentialEquation}) for $i=0\sim MN/(2n)-1$, then
$\{key^*(i)\}_{i=0}^{MN/(2n)-1}$ can be used as an equivalent key to decrypt any cipher-images of smaller size, encrypted with the same secret key.

The complexity of the differential attack is mainly determined by verifying the $n-1$ bits of each element in $\{key^*(i)\}_{i=0}^{MN/(2n)-1}$ from
Table~\ref{table:carrybits}, so the complexity is proportional to the size of the plain-image.

\subsection{Breaking the Secret Key}

The differential attack described in the above subsection only outputs an equivalent key, which can only be used to decrypt other cipher-images of smaller size than that of the chosen plain-images. To decrypt any other cipher-image encrypted with the same secret key, we need to obtain the secret key. How to derive it from the equivalent key will be discussed in this sub-section.

Assume $\{b(l)\}$ distributes over $\{0, 1\}$ uniformly, the probability $key_1 \not\in (key^*(i))_{i=0}^{8MN/n-1}$ or
$key_2 \not\in (key^*(i))_{i=0}^{8MN/n-1}$ is $(1/2)^{8MN/n}$. So, we can obtain set $(key_1, key_2)$ with a very high probability
$1-(1/2)^{8MN/n-1}$. Since $\sum_{j=0}^{n-2}(key_{1,j}\cdot 2^j)\neq \sum_{j=0}^{n-2}(key_{2,j}\cdot 2^j)$, one can narrow the scope of $B(i)$ from Eq.~(\ref{eq:DifferentialEquation})
as follows
\begin{equation}
B(i)\in
\begin{cases}
\{2,3\},  & \mbox{if } \sum_{j=0}^{n-2}(key^*(i)_j\cdot 2^j)=\sum_{j=0}^{n-2}(key_{1,j}\cdot 2^j),\\
\{0,1\},  & \mbox{if } \sum_{j=0}^{n-2}(key^*(i)_j\cdot 2^j)=\sum_{j=0}^{n-2}(key_{2,j}\cdot 2^j),
\end{cases}
\label{eq:determinebits1}
\end{equation}
where $key^*(i)=\sum_{j=0}^{n-1}(key^*(i)_j\cdot 2^j)$ and $key^*(i)=\sum_{j=0}^{n-1}(key^*(i)_j\cdot 2^j)$.

\begin{proposition}
Assume that $a$ and $x$ are both $n$-bit integers, $n\in \mathbb{Z}^+$, if $a$ is odd, then
$p=((a\dotplus x)\oplus x)$ is always odd and $q=((a\dotplus x)\odot x)$ is always even.
\label{propsition:lsb}
\end{proposition}
\begin{proof}
This proposition can be proved by two equations
\begin{eqnarray*}
((1+x_0)\bmod 2)\oplus x_0 & \equiv & 1,\\
((1+x_0)\bmod 2)\odot x_0 &  \equiv & 0.
\end{eqnarray*}
\end{proof}

From Proposition~\ref{propsition:lsb} and Eq.~(\ref{eq:Encrypt}), one can narrow the scope of $B(i)$ also according to encryption result
of the second chosen intermediate sequence shown in Eq.~(\ref{eq:scondchooseimage}), as follows
\begin{equation}
B(i)\in
\begin{cases}
\{1, 3\},  & \mbox{if $J_1'(i)$ is odd},\\
\{0, 2\},  & \mbox{if $J_1'(i)$ is even}.
\end{cases}
\label{eq:determinebits2}
\end{equation}

Once $key_1$ and $key_2$ are determined, value of $B(i)$, for $i=0\sim 8MN/n-1$, can be determined exactly from Eq.~(\ref{eq:determinebits1}) and Eq.~(\ref{eq:determinebits2}).
There are only two possible combinations of $key_1$ and $key_2$. If the searched version is the right one, $\{B(i)\}_{i=0}^{8MN/n-1}$ can be constructed correctly.
Let $\{B^{\star}(i)\}_{i=0}^{8MN/n-1}$ and $\{B^*(i)\}_{i=0}^{8MN/n-1}$ denote the obtained version of $\{B(i)\}_{i=0}^{8MN/n-1}$ corresponding to the two combinations of $key_1$ and $key_2$.
Since Eq.~(\ref{eq:determinebits2}) is unrelated with $key_1$ and $key_2$, one can assure that $B^{\star}(i)=B^{*}(i)\oplus 2$, i.e, $b^{\star}(2i)=1-b^*(2i)$ and $b^{\star}(2i+1)=b^*(2i+1)$, for $i=0\sim 8MN/n-1$.
Construct $\{x^{\star}(i)\}_{i=0}^{MN/(2n)-1}$ and $\{x^*(i)\}_{i=0}^{MN/(2n)-1}$, where $x^{\star}(i)=\sum_{j=1}^{32}b^{\star}(32\cdot i+j-1)\cdot 2^{-j}$, $x^*(i)=\sum_{j=1}^{32}b^*(32\cdot i+j-1)\cdot 2^{-j}$.

Since $\{x(i)\}_{i=0}^{MN/(2n)-1}$ come from consecutive chaotic states generated by iterating Logistic map, we can distinguish $\{x^{\star}(i)\}_{i=0}^{MN/(2n)-1}$ or $\{x^*(i)\}_{i=0}^{MN/(2n)-1}$
is the right sequence controlling encryption process, and verify $key_1$ and $key_2$ correspondingly, by checking whether any two consecutive elements of them satisfy specific correlation. As shown in \cite[Table 3]{Rao:ModifiedCKBA:ICDSP07}, Eq.~(\ref{equation:Logistic}) is realized in 32-bit fixed-point arithmetic precision. So MCKBA satisfies condition described in Proposition~\ref{propsition:muerror} with $L=32$.
The whole secret key of MCKBA can be verified by checking whether some consecutive elements in $\{x^{\star}(i)\}_{i=0}^{MN/(2n)-1}$ and $\{x^*(i)\}_{i=0}^{MN/(2n)-1}$ satisfy Eq.~(\ref{eq:muerror}).
Finally, $key_1$, $key_2$, and $x(0)=\sum_{j=1}^{32}b(j-1)\cdot 2^j$ can be recovered. For HCKBA, we have to check which sequence agrees with distribution of the chaotic states generated by the hyperchaos generator like \cite[Fig.~6]{CH:HCKBA:IJBC10}.

\begin{proposition}
Assume that the Logistic map $x(k+1)=\mu\cdot x(k)\cdot (1-x(k))$ is iterated with $L$-bit fixed-point
arithmetic and that $x(k+1)\ge 2^{-m}$, where $1\le m\le L$. Then, the following inequality holds
\begin{equation}
|\mu-\tilde{\mu}_k|\le 2^{m+3}/2^L,
\label{eq:muerror}
\end{equation}
where $\tilde{\mu}_k=\frac{x(k+1)}{x(k)\cdot (1-x(k))}$.
\label{propsition:muerror}
\end{proposition}
\begin{proof}
See appendix of \cite{Li:AttackingRCES2008}.
\end{proof}

\subsection{Experimental Results}

\begin{figure}[!htb]
\centering
\begin{minipage}{\figwidth}
\centering
\includegraphics[width=\textwidth]{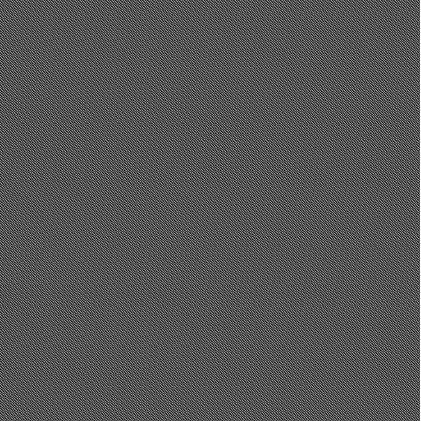}
a)
\end{minipage}
\begin{minipage}{\figwidth}
\centering
\includegraphics[width=\textwidth]{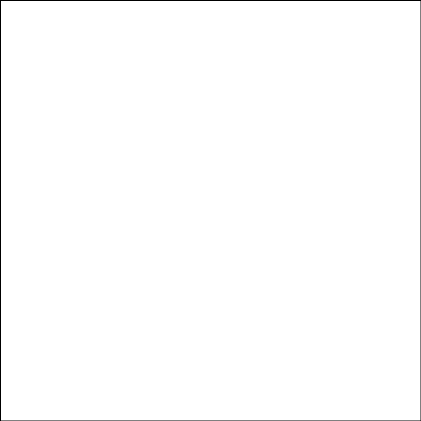}
b)
\end{minipage}\\
\begin{minipage}{\figwidth}
\centering
\includegraphics[width=\textwidth]{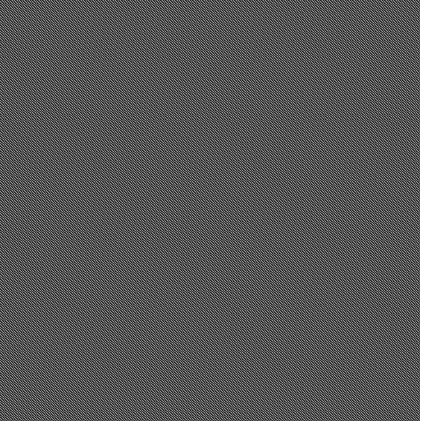}
c)
\end{minipage}
\begin{minipage}{\figwidth}
\centering
\includegraphics[width=\textwidth]{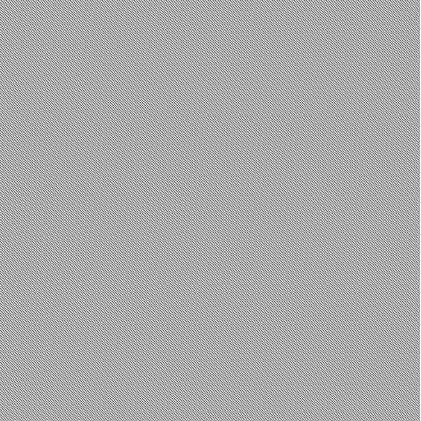}
d)
\end{minipage}
\caption{Four chosen plain-images (The black boundary of Fig.~\ref{fig:ChosenImages}b) is not its part).}
\label{fig:ChosenImages}
\end{figure}

\begin{figure}[!htb]
\centering
\begin{minipage}{\figwidth}
\centering
\includegraphics[width=\textwidth]{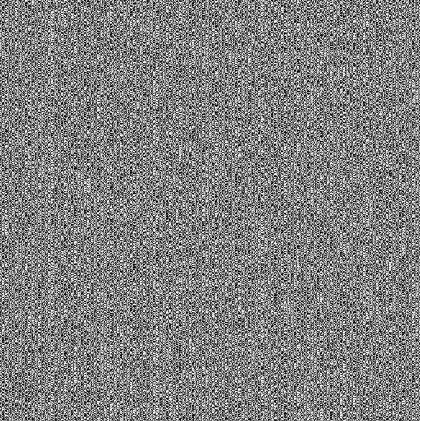}
a)
\end{minipage}
\begin{minipage}{\figwidth}
\centering
\includegraphics[width=\textwidth]{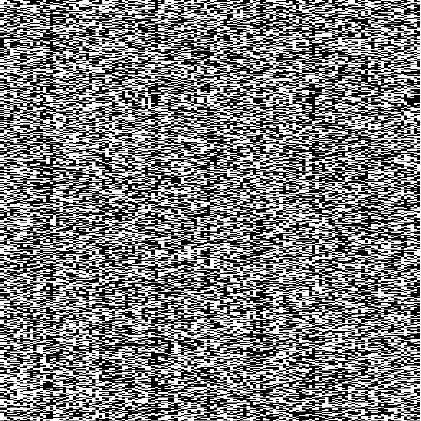}
b)
\end{minipage}\\
\begin{minipage}{\figwidth}
\centering
\includegraphics[width=\textwidth]{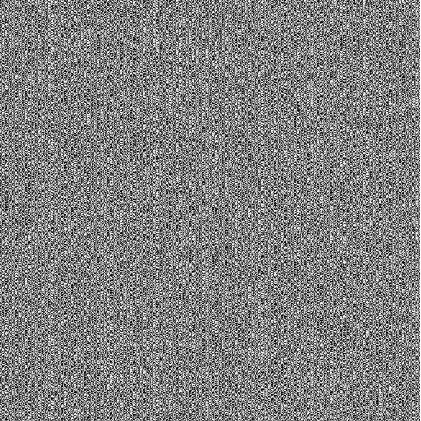}
c)
\end{minipage}
\begin{minipage}{\figwidth}
\centering
\includegraphics[width=\textwidth]{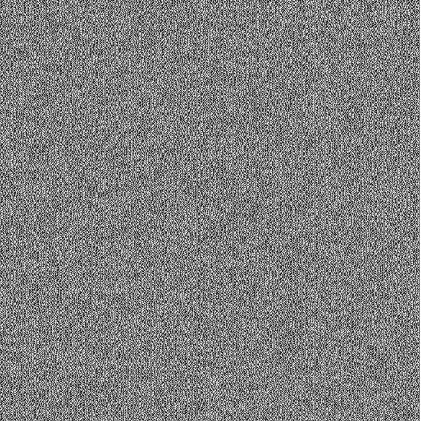}
d)
\end{minipage}
\caption{The corresponding cipher-images of the four chosen plain-images shown in Fig.~\ref{fig:ChosenImages}.}
\label{fig:EncryptedChosenImages}
\end{figure}

\begin{figure}[!htb]
\centering
\begin{minipage}{\figwidth}
\centering
\includegraphics[width=\textwidth]{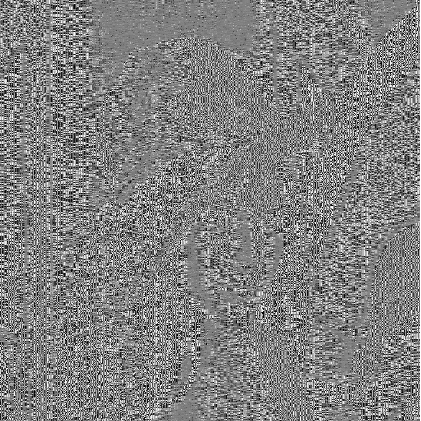}
a)
\end{minipage}
\begin{minipage}{\figwidth}
\centering
\includegraphics[width=\textwidth]{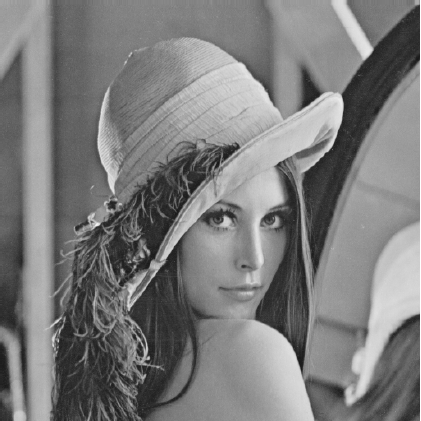}
b)
\end{minipage}
\caption{The decryption result of another cipher-image encrypted with the same secret key: a) cipher-image; b)
decrypted plain-image.}
\label{fig:DecryptedLenna}
\end{figure}

To verify the real performance of the above analysis, some experiments are carried out on some plain-images of size $512\times 512$ when $n=32$. The four chosen
plain-images are shown in Fig.~\ref{fig:ChosenImages}. When $x_0=319684607/2^{32}$, $key_1=3835288501$, and $key_2=1437224678$, the encryption results of the four
chosen-image are shown in Fig.~\ref{fig:EncryptedChosenImages}. Equivalent key $\{key^*(i)\}_{i=0}^{MN/(2n)-1}$ is used to decrypt another cipher-image
shown in Fig.~\ref{fig:DecryptedLenna}a) and the recovered result is shown in Fig.~\ref{fig:DecryptedLenna}b). In this case, the three parts of the whole secret key,
$key_1$, $key_2$ and the 32 bits of $x(0)$ can be verified by checking only one pair of consecutive elements in $\{x^{\star}(i)\}_{i=0}^{MN/(2n)-1}$ and $\{x^*(i)\}_{i=0}^{MN/(2n)-1}$.

\subsection{Some Remarks on the Performance of MCKBA}

\begin{itemlist}
\item Insufficient randomness of PRBS $\{b(l)\}$

It is well-known that distribution of chaotic states generated by iterating Logistic map is not uniform, which
makes randomness of derived binary bit sequence from them very low. As this point has been shown quantitatively in
\cite{Li:AttackingBitshiftXOR2007,LiLi:BreakPareek:CNSNS10}, detailed discussion is omitted here.

\item Insensitivity with respect to changes of plain-image

This defect may cause serious threat for any secure image encryption algorithm since image and its watermarked version
may be encrypted at the same time. From Eq.~(\ref{eq:Encrypt}), one can see that change of the $m$-th significant bit of $J(i)$
may only change the $m\sim (n-1)$-th significant bits of $J'(i)$, where $0\le m<n$. This means MCKBA can make
change of one bit of plain-image to influence at most $n$ bits in the corresponding cipher-image.

\item Insensitivity with respect to changes of two sub-keys

Obviously, any secure encryption algorithm should avoid this defect. Unfortunately, MCKBA is  seriously fragile in this aspect.
From Eq.~(\ref{eq:Encrypt}), one can see that change of the $m$-th significant bit of $key_1$ or $key_2$ only influences
the $m\sim (n-1)$-th significant bits of the corresponding cipher-pixel. As shown in Proposition~\ref{propsition:msb}, change of
the most significant bit of $key_1$ or $key_2$ has no any influence on the whole decryption.
\end{itemlist}

\section{Conclusion}

In this paper, security of the image encryption algorithm MCKBA has been studied in detail.
It was found that the whole secret key can be recovered correctly with only four chosen plain-images.
In addition, some other defects of the algorithm, including insensitivity with respect to changes of
plain-image/secret key, were discussed. Analogue of MCKBA, HCKBA, has the same security problems.
Due to such a low level of security provided by the two algorithms (essentially one algorithm),
their application in practice should be performed with extreme caution.

\section*{Acknowledgement}

The work of Chengqing Li was partially supported by The Hong Kong Polytechnic
University's Postdoctoral Fellowships Scheme under grant no. G-YX2L.

\bibliographystyle{ws-ijbc}
\bibliography{MCKBA}

\begin{thebibliography}{27}
\newcommand{\enquote}[1]{``#1''}
\providecommand{\natexlab}[1]{#1}
\providecommand{\url}[1]{\texttt{#1}}
\providecommand{\urlprefix}{URL }
\expandafter\ifx\csname urlstyle\endcsname\relax
  \providecommand{\doi}[1]{doi:\discretionary{}{}{}#1}\else
  \providecommand{\doi}{doi:\discretionary{}{}{}\begingroup
  \urlstyle{rm}\Url}\fi

\bibitem[{\'{A}lvarez \& Li(2006)}]{AlvarezLi:Rules:IJBC2006}
\'{A}lvarez, G. \& Li, S. [2006] \enquote{Some basic cryptographic requirements
  for chaos-based cryptosystems,} \emph{International Journal of Bifurcation
  and Chaos} \textbf{16},  2129--2151.

\bibitem[{Arroyo \emph{et~al.}(2008)Arroyo, Rhouma, Alvarez, Li \&
  Fernandez}]{David:AttackingChaos08}
Arroyo, D., Rhouma, R., Alvarez, G., Li, S. \& Fernandez, V. [2008] \enquote{On
  the security of a new image encryption scheme based on chaotic map lattices,}
  \emph{Chaos} \textbf{18},  art. no. 033112.

\bibitem[{Chen \emph{et~al.}(2004)Chen, Mao \& Chui}]{YaobinMao:CSF2004}
Chen, G., Mao, Y. \& Chui, C.~K. [2004] \enquote{A symmetric image encryption
  scheme based on 3{D} chaotic cat maps,} \emph{Chaos, Solitons \textup{\&}
  Fractals} \textbf{21},  749--761.

\bibitem[{Chen \& Yen(2003)}]{Chen&Yen:RCES:JSA2003}
Chen, H.-C. \& Yen, J.-C. [2003] \enquote{A new cryptography system and its
  {VLSI} realization,} \emph{Journal of Systems Architecture} \textbf{49},
  355--367.

\bibitem[{Gangadhar \& Rao(2010)}]{CH:HCKBA:IJBC10}
Gangadhar, C. \& Rao, K.~D. [2010] \enquote{Hyperchaos based image encryption,}
  \emph{International Journal of Bifurcation and Chaos} \textbf{19},
  3833--3839.

\bibitem[{Li \emph{et~al.}(2007)Li, Li, \'{A}lvarez, Chen \&
  Lo}]{Li:AttackingBitshiftXOR2007}
Li, C., Li, S., \'{A}lvarez, G., Chen, G. \& Lo, K.-T. [2007]
  \enquote{Cryptanalysis of two chaotic encryption schemes based on circular
  bit shift and {XOR} operations,} \emph{Physics Letters A} \textbf{369},
  23--30.

\bibitem[{Li \emph{et~al.}(2009{\natexlab{a}})Li, Li, Asim, Nunez, Alvarez \&
  Chen}]{Li:AttackingIVC2009}
Li, C., Li, S., Asim, M., Nunez, J., Alvarez, G. \& Chen, G.
  [2009{\natexlab{a}}] \enquote{On the security defects of an image encryption
  scheme,} \emph{Image and Vision Computing} \textbf{27},  1371--1381.

\bibitem[{Li \emph{et~al.}(2009{\natexlab{b}})Li, Li, Chen \&
  Halang}]{Li:BreakImageCipher:IVC09}
Li, C., Li, S., Chen, G. \& Halang, W.~A. [2009{\natexlab{b}}]
  \enquote{Cryptanalysis of an image encryption scheme based on a compound
  chaotic sequence,} \emph{Image and Vision Computing} \textbf{27},
  1035--1039.

\bibitem[{Li \emph{et~al.}(2004{\natexlab{a}})Li, Li \&
  Lo}]{LiLi:BreakPareek:CNSNS10}
Li, C., Li, S. \& Lo, K.-T. [2004{\natexlab{a}}] \enquote{Breaking a modified
  substitution-diffusion image cipher based on chaotic standard and logistic
  maps,} \emph{Communications in Nonlinear Science and Numerical Simulation}
  \textbf{16},  837--843.

\bibitem[{Li \emph{et~al.}(2004{\natexlab{b}})Li, Chen \&
  Zheng}]{Li:ChaosImageVideoEncryption:Handbook2004}
Li, S., Chen, G. \& Zheng, X. [2004{\natexlab{b}}] ``4,'' \emph{Chaos-Based
  Encryption for Digital Images and Videos}, Multimedia Security Handbook (CRC
  Press), pp. 133--167.

\bibitem[{Li \emph{et~al.}(2008{\natexlab{a}})Li, Li, Chen, Bourbakis \&
  Lo}]{Li:AttackingPOMC2008}
Li, S., Li, C., Chen, G., Bourbakis, N.~G. \& Lo, K.-T. [2008{\natexlab{a}}]
  \enquote{A general quantitative cryptanalysis of permutation-only multimedia
  ciphers against plaintext attacks,} \emph{Signal Processing: Image
  Communication} \textbf{23},  212--223.

\bibitem[{Li \emph{et~al.}(2008{\natexlab{b}})Li, Li, Chen \&
  Lo}]{Li:AttackingRCES2008}
Li, S., Li, C., Chen, G. \& Lo, K.-T. [2008{\natexlab{b}}]
  \enquote{Cryptanalysis of the {RCES/RSES} image encryption scheme,}
  \emph{Journal of Systems and Software} \textbf{81},  1130--1143.

\bibitem[{Li \& Zheng(2002)}]{Li-Zheng:CKBA:ISCAS2002}
Li, S. \& Zheng, X. [2002] \enquote{Cryptanalysis of a chaotic image encryption
  method,}  \emph{Proceedings of IEEE International Symposium on Circuits and
  Systems}, pp. 708--711.

\bibitem[{Pisarchik \emph{et~al.}(2006)Pisarchik, Flores-Carmona \&
  Carpio-Valadez}]{Flores:EncryptLatticeChaos06}
Pisarchik, A.~N., Flores-Carmona, N.~J. \& Carpio-Valadez, M. [2006]
  \enquote{Encryption and decryption of images with chaotic map lattices,}
  \emph{Chaos} \textbf{16},  art. no. 033118.

\bibitem[{Rao \& Gangadhar(2007)}]{Rao:ModifiedCKBA:ICDSP07}
Rao, K. \& Gangadhar, C. [2007] \enquote{Modified chaotic key-based algorithm
  for image encryption and its {VLSI} realization,}  \emph{Proceedings of the
  2007 15th International Conference on Digital Signal Processing}, pp.
  439--442.

\bibitem[{Rhouma \& Belghith(2008)}]{Rhouma:BreakLian:PLA08}
Rhouma, R. \& Belghith, S. [2008] \enquote{Cryptanalysis of a spatiotemporal
  chaotic image/video cryptosystem,} \emph{Physics Letters A} \textbf{372},
  5790--5794.

\bibitem[{Socek \emph{et~al.}(2005)Socek, Li, Magliveras \&
  Furht}]{SocekLi:SecureComm2005}
Socek, D., Li, S., Magliveras, S.~S. \& Furht, B. [2005] \enquote{Enhanced
  {1-D} chaotic key-based algorithm for image encryption,}  \emph{Proceedings
  of the First IEEE/CreateNet International Conference on Security and Privacy
  for Emerging Areas in Communication Networks (SecureComm 2005)}, pp.
  406--408.

\bibitem[{Solak \& Cokal(2009)}]{Solak:BreakXiang:PLA09}
Solak, E. \& Cokal, C. [2009] \enquote{Algebraic break of a cryptosystem based
  on discretized two-dimensional chaotic maps,} \emph{Physics Letters A}
  \textbf{373},  1352--1356.

\bibitem[{Solak \emph{et~al.}(2010)Solak, Cokal, Yildiz \&
  Biyikoglu}]{Solak:Fridrich:IJBC10}
Solak, E., Cokal, C., Yildiz, O.~T. \& Biyikoglu, T. [2010]
  \enquote{Cryptanalysis of {F}ridrich's chaotic image encryption,}
  \emph{International Journal of Bifurcation and Chaos} \textbf{20},
  1405--1413.

\bibitem[{Takahashi \emph{et~al.}(2004)Takahashi, Nakano \&
  Saito}]{Takahashi:HyperChaos:TCASII04}
Takahashi, Y., Nakano, H. \& Saito, T. [2004] \enquote{A simple hyperchaos
  generator based on impulsive switching,} \emph{IEEE Transactions on Circuits
  and Systems II-Express Briefs} \textbf{51},  468--472.

\bibitem[{Wang \emph{et~al.}(2005)Wang, Pei, Zou, Song \& He}]{Kaiwang:PLA2005}
Wang, K., Pei, W., Zou, L., Song, A. \& He, Z. [2005] \enquote{On the security
  of 3{D} cat map based symmetric image encryption scheme,} \emph{Physics
  Letters A} \textbf{343},  432--439.

\bibitem[{Wong \emph{et~al.}(2010)Wong, Lin \&
  Chen}]{KWong:chaoticoding:TCASSII10}
Wong, K.-W., Lin, Q. \& Chen, J. [2010] \enquote{Simultaneous arithmetic coding
  and encryption using chaotic maps,} \emph{IEEE Transactions on Circuits and
  Systems II-Express Briefs} \textbf{57},  146--150.

\bibitem[{Xiang \emph{et~al.}(2007)Xiang, Wong \& Liao}]{Xiang:Encrypt:PLA07}
Xiang, T., Wong, K.-W. \& Liao, X. [2007] \enquote{A novel symmetrical
  cryptosystem based on discretized two-dimensional chaotic map,} \emph{Physics
  Letters A} \textbf{364},  252--258.

\bibitem[{Yang \emph{et~al.}(2011)Yang, Xiao \&
  Xiang}]{YangXiao:4RBlock:CNSNS10}
Yang, J., Xiao, D. \& Xiang, T. [2011] \enquote{Cryptanalysis of a chaos block
  cipher for wireless sensor network,} \emph{Communications in Nonlinear
  Science and Numerical Simulation} \textbf{16},  844--850.

\bibitem[{Ye(2010)}]{Ye:Scramble:PRL10}
Ye, G. [2010] \enquote{Image scrambling encryption algorithm of pixel bit based
  on chaos map,} \emph{Pattern Recognition Letters} \textbf{31},  347--354.

\bibitem[{Yen \& Guo(2000)}]{Yen:CKBA:ISCAS2000}
Yen, J.-C. \& Guo, J.-I. [2000] \enquote{A new chaotic key-based design for
  image encryption and decryption,}  \emph{Proceedings of IEEE International
  Symposium on Circuits and Systems}, pp. 49--52.

\bibitem[{Zhou \& Au(2008)}]{Zhou:CommentChaoticrypt:TCASI08}
Zhou, J. \& Au, O.~C. [2008] \enquote{Comments on ``a novel compression and
  encryption scheme using variable model arithmetic coding and coupled chaotic
  system",} \emph{IEEE Transactions on Circuits and Systems I} \textbf{55},
  3368--3369.

\end{thebibliography}
\end{document}